\documentstyle[preprint,aps]{revtex}
\begin{document}
\draft
\tightenlines
\title
{\bf Low--energy photodisintegration of $^9$Be and $\alpha +\alpha + n
\leftrightarrow ^9$Be$+\gamma$\\
reactions at astrophysical conditions}
\author{V.D. Efros$^{1)}$, H. Oberhummer$^{2)}$, A. Pushkin$^{3)}$, and
I.J. Thompson$^{4)}$}
\address{
1) Russian Research Centre ``Kurchatov Institute'', Kurchatov Square 1,
123182 Moscow, Russia\\
2) Institut f\"ur Kernphysik, Wiedner Hauptstrasse 8--10, TU Wien, A--1040, Vienna, Austria\\
3) Department of Mathematical Physics, Lund Institute of Technology,\\ P.O. Box 118 S--221 00 Lund, Sweden\\ 
4) Department of Physics, University of Surrey, Guildford GU2 5XH, United Kingdom
}

\date{\today}
\maketitle
\begin{abstract}
A semi--microscopic model for the low--energy photodisintegration
of the $^9$Be nucleus is constructed, and the experimental data are analyzed with its help.  
The older radioactive isotope data are supported by this analysis. The theoretical
photodisintegration 
cross section
is derived. The astrophysical rates
for the reaction $ \alpha+\alpha+n\rightarrow ^9$Be$+\gamma$ and the reverse
photodisintegration of $^9$Be
are calculated. The new reaction rate for $\alpha+\alpha+n\rightarrow ^9$Be$+\gamma$
is compared with previous estimations.
%At higher
%temperatures the rates obtained are several times larger than those found previously.
\end{abstract}
\bigskip
\pacs{PACS numbers: 25.20.Dc, 25.40.Lw, 97.10Cv, 21.60.Gx 21.45.+v,\\}

\section{INTRODUCTION}
Recently fully microscopic calculations of nuclei with $A\le 9$ have become
feasible \cite{pudl97,varga96}. 
The $^9$Be nucleus is such a system of special interest, as it allows 
tests of theories of interaction of composite particles \cite{varga96}. 
The properties of low--energy continuum of $^9$Be are of
particular importance in this connection.
However, the corresponding experimental data on the low--energy 
photodisintegration
of $^9$Be are not in mutual agreement (see Fig.~1).
In the present work we develop a semi--microscopic model to describe the 
process,
and we analyze the experimental data
with its help. The model accounts simultaneously for both resonant and non--resonant
contributions to the cross section. An estimation of the 
reliability of various data is obtained and a theoretical
photodisintegration cross section is derived.

We also calculate the reaction rates of the reaction
$^9$Be$+\gamma\rightarrow\alpha+\alpha+n $ and the reverse reaction
for astrophysical conditions. These reaction rates are of relevance in
the high--entropy bubble in type II supernovae,
an astrophysical site that has been suggested for the r--process
\cite{woo94,tak94}. The baryonic matter in this bubble is dominated
in the beginning by $\alpha$--particles, neutrons, and
protons. The abundance distribution shifts then to higher masses
through the recombination of the free $\alpha$--particles, neutrons, and
protons. This generates the so--called $\alpha$--process leading to the formation
of massive isotopes ($A \approx 100$). The reaction path
in the $\alpha$--process is mainly determined
by requirements of nuclear statistical equilibrium and depends also on
the reaction rates of the various recombination paths bridging
the mass 5 and 8 gaps. It has been shown that
there are three principal reaction paths from $^4$He to $^{12}$C:\\
(i) $^4$He(2$\alpha$,$\gamma$)$^{12}$C\\
(ii) $^4$He($\alpha$ n,$\gamma$)$^9$Be($\alpha$,n)$^{12}$C\\
(iii) $^4$He(2n,$\gamma$)$^6$He($\alpha$,n)$^9$Be($\alpha$,n)$^{12}$C.

It was shown in Refs.~\cite{tak94,woo92,gor95} that the
triple--alpha process (i) can be neglected compared
to the reaction sequence (ii) via $^9$Be
under r--process conditions in the $\alpha$--process.
Also the reaction path (iii) via
$^6$He can be neglected for this scenario
\cite{gor95,efr96}. This is true even if the
reaction rate of $^4$He(2n,$\gamma$)$^6$He is
strongly enhanced~\cite{her97}, because then $^6$He is also destroyed very effectively
through photodissociation. Therefore, for the 
$\alpha$-- and r--process the reaction
$^4$He($\alpha$ n,$\gamma$)$^9$Be
plays a key role in bridging the unstable mass gaps
at $A=5$ and $A=8$. 

The reaction rates of $^4$He($\alpha$ n,$\gamma$)$^9$Be
and the reverse photodisintegration of $^9$Be were estimated in Ref.~\cite{fowler75}
from the experimental photodisintegration cross section.
However, Ref.~\cite{fowler75} did not include information on which
experimental data their estimate was based. In view of the
astrophysical relevance of these reactions we recalculate in the present work 
the rates of the first step of the reaction (ii) above.

The same problem is also addressed in Ref.~\cite{gor95}. These authors obtain
the resonant contribution to the
$^9$Be$(\gamma,n)^8$Be cross section from
the Breit--Wigner formula for the first excited state of
$^9$Be with the parameters taken from Ref.~\cite{ajz88}.\footnote{We note that
the $\Gamma$ and $E_0$ parameters of the resonance used in
Ref.~\cite{gor95} seem to be incorrect. 
The resonant properties of the $1/2^+$ state of $^9$Be
will be considered in our future work.}
In order to calculate the non--resonant contribution they
introduce a single--particle potential with the depth chosen
to reproduce the ground state, calculate both ground-- and final--state 
continuum wave functions in this potential, and multiply the cross section obtained 
by the shell--model spectroscopic factor. They then add this cross section
constructively or destructively to the resonant cross section to
establish possible upper and lower bounds for the reaction rates.
This procedure has certain shortcomings: a resonant
contribution to the cross section should not emerge
as an addition to the dynamic model used, since a correct
quantum mechanical model should necessarily contain such a contribution itself,
along with the non--resonant contribution and an interference term.
Besides, the potential wells used for the
ground state and continuum state should in fact be different:
an additional spin--orbit potential, for example, should be present in the
ground p--state as compared to the continuum s--state.

In our model we use a three-body specification of the $^9$Be bound state,
and a semimicroscopic continuum wave function which describes the essential 
scattering degrees of freedom at low relative energies.
In Sect.~2 this model is formulated. In Sect.~3 the 
results for the $^9$Be$(\gamma,n)$ cross section
are given.
In Sect.~4 the astrophysical rates for
$\alpha+\alpha+n\rightarrow ^9$Be$+\gamma$ and the reverse reaction
are calculated.

\section{The model}

The relevant experimental data on the low--energy
$^9$Be$(\gamma,n)$ cross
section are presented in Fig.~1. The available data are those in 
\cite{jacob61,berman67} obtained with
bremsstrahlung photons and those in
\cite{gib59,john62,ham53,fui82} obtained from $\gamma$--radiation
from radioactive isotopes. The peak at very
low energy exhibited by the data of Ref. \cite{berman67} is not
confirmed by other groups, and may arise
from discrepancies caused by neutron energy loss in the targets \cite{bar68}.
The radioactive isotope techniques
normally provide more reliable results due to the absence of
difficulties with the energy resolution. However, the cross
section can be determined only for a few discrete photon
energies with this method. This drawback will be cured below
by the use of an appropriate theoretical
model. Our strategy will be thus to analyze the
radioactive isotope data. 
We shall consider the range of energies up to 0.5\,MeV above
threshold.

We need to obtain the ground state and continuum wave functions (WF)
and calculate the transition matrix element.
We start with the three--body $\alpha+\alpha+n$ representation
of the $^9$Be system. Within this representation the WF
in the c.m. system is the $\alpha\alpha n$ relative
motion function times the intrinsic WFs
of the two $\alpha$--particles. Since a predominant
contribution to the transition matrix element comes from
distances large compared to the size of the $\alpha$--particle
additional antisymmetrization may be disregarded.\footnote{The
minor role of the antisymmetrization is seen
from the calculations of the $\alpha+n+p$ system \cite{kukulin}, for example.}
The intrinsic
$\alpha$ particle WFs then will drop out from the
calculation. In the following we shall refer to the
three--particle relative motion function as to the WF
of the system.

Let us denote \mbox{\boldmath $\rho$} and $\bf r$ the distance between the
$\alpha$ particles and that from their center of mass to the
neutron, respectively. The ground state is $J^\pi=3/2^-$, and
its WF is of the form
\begin{equation}
\Psi_{\rm gs}=\sum_{l_1l_2L}\phi_{l_1l_2L}(\rho,r)
[[Y_{l_1}(\hat{\rho})Y_{l_2}(\hat{r})]_L\chi_{1/2}]_J.
\label{eq:gswf}
\end{equation}
Here $\chi$ is the neutron spin function, and the brackets
$[\ldots]$ stand for angular momentum coupling. Because of the Pauli principle
and parity requirements $l_1$ is even, and $l_2$ is odd.
The WF in Eq.~(\ref{eq:gswf}) was obtained from the three--body
Schr\"odinger
equation with $\alpha\alpha$ and $\alpha n$ potentials reproducing
the observed two--body phase shifts.
These potentials, along with some details of the calculation, are listed
in the appendix. Practically an exact solution to the
$\alpha+\alpha+n$ bound state problem is obtained, but
one cannot get the experimental binding
energy with these ``bare'' interparticle interactions. A possible reason is 
that the two $\alpha$--particles may distort each other in the 
$\alpha+\alpha+n$ bound state compared to the pure $n\alpha$ case. This may lead
to a change in the $n\alpha$ interaction.
To obtain a
reasonable ground state WF the strength of the attractive central
component of the $n\alpha$ potential is reduced by 8\% in our
calculation. This
leads to values of 1.50\,MeV and 2.48\,fm for the binding
energy and charge radius of $^9$Be, sufficiently close to the
experimental values of 1.5736\,MeV and ($2.51\pm 0.01$)\,fm.

Coming to the continuum wave function, we note first
that the $^8$Be resonance produced in the reaction may be
safely treated as a stable particle for the purposes of our calculation. This is
because its width of 7\,keV is extremely narrow on a nuclear
scale. The $\alpha\alpha$ continuum wave function, taken at the
resonance energy and normalized to unity in the interior region,
decreases practically to zero in the Coulomb barrier region as shown in Fig.~2 (solid line).
This function represents the WF of the resonance extremely well and
is taken as the $^8$Be ``bound state'' wave function.
Second, we argue that photodisintegration of $^9$Be
proceeds entirely into the $^8$Be$+n$ channel. Indeed, one can estimate that, at small
energies considered, the three--fragment $\alpha+\alpha+n$ disintegration channel is strongly
suppressed due to the threshold regime. The experimental data also strongly supports
the absence of this channel \cite{berman67}. At the same time,
the two--fragment $^4$He$+^5$He channel is still closed and ineffective due
to the broad width of $^5$He. Thus our cross section starts
from the $^8$Be$+n$ threshold of 1.6654\,MeV. The cross section
in the region between this threshold and the $\alpha\alpha n$
threshold of 1.5736\,MeV is known to be tiny \cite{fui83} and will be disregarded.
As in the previous work (e.g. \cite{francis60}) we confine
ourselves to an s--wave relative $n^8$Be motion, i.e., with
$1/2^+$ continuum states.

Predictions of the above dynamic $\alpha\alpha n$ model for the
photodisintegration of $^9$Be
depend crucially on the position of the excited state of
$^9$Be with respect to the threshold. Preliminary three--body
calculations gave us a peak lying too
high in energy, and too broad. This could not be used for
a reasonable fit to the data, so in the following we shall
formulate an alternative representation of the continuum WF. We shall
seek it in the form
\begin{equation}
\Psi_{\rm f}=\frac{1}{\sqrt{4\pi}}\frac{\phi^{^8{\rm Be}}(\rho)}{\rho}
\frac{1}{\sqrt{4\pi}}
\frac{\psi_{\rm sc}(r)}{r}\chi_{1/2}.   \label{eq:cwf}
\end{equation}
Here $\phi^{^8{\rm Be}}$ is the intrinsic wave function of $^8$Be
calculated with the same $\alpha\alpha$ potential as for the
ground state, and $\psi_{\rm sc}$ is the $n^8$Be relative motion
function, where for large $r$ the normalization
\begin{equation}
\psi_{\rm sc}(r)\rightarrow \sin(kr+\delta) \label{eq:sin}
\end{equation}
is used.

Generally speaking, the true continuum wave function
differs from Eq. (\ref{eq:cwf}) not only in the $n^8$Be interaction
region but also in the outer region. However, at energies in the
vicinity of the long--living excited state of $^9$Be the
representation (\ref{eq:cwf}) should approximately be valid in the
outer region. Indeed, the decay of the long--living state into the three--body
$\alpha\alpha n$ channel is inhibited due to the threshold regime.
Due to the approximate validity of the WF in Eq.~(\ref{eq:cwf})
in the outer region one obtains 
a correct energy dependence of the cross section
when using this WF. For small energies the main 
energy dependence of the transition matrix element appears as a factor 
in the continuum WF and is determined by an outer part of the WF, i.e.
the phase shift. Besides,
one can see below that just the outer region (where $\psi_{\rm sc}$ takes
the form of Eq.~(\ref{eq:sin})) gives the biggest contribution
to the transition matrix element. 
One can therefore hope that a WF of the form of Eq. (\ref{eq:cwf}) 
suffices for the fitting purpose in the whole energy
range considered.

We seek the relative motion function $\psi_{\rm sc}$ as a
solution to the
relative motion Schr\"odinger equation with some potential whose
parameters are chosen from a fit of the theoretical cross
section to the data. Taking into account that the s--wave
$\alpha n$ repulsion
and the p--wave $\alpha n$ attraction have comparable ranges
one can assume a smooth attractive potential. The Woods--Saxon
family
\begin{equation}
V(r)=-V_0\left(1+e^{\frac{r-R}{a}}\right)^{-1} \label{eq:ws}
\end{equation}
will be adopted below as a good representative.

%Attractive potentials
%could lead to the description of the first excited level of $^9$Be
%as a virtual state or a bound state (bound within the $^8$Be$+n$
%model) but not as a resonant state that would require a strong
%potential
%barrier for the s--state neutron. While some $R$ matrix fits to
%the data lead to
%a $^8$Be$+n$ resonant state, they seem to be rather artificial due to
%the above reason.

Consider now the representation of the photodisintegration
cross section in our model. The matrix element of the
dipole transition operator
$(4/9)e{\bf r}$ between the WFs in Eq.~(\ref{eq:gswf}) and Eq.~(\ref{eq:cwf})
has to be calculated. After integrating over $\bf \rho$
the matrix element reduces to the overlap between the $n^8$Be
relative motion functions, namely the scattering function entering
Eq. (\ref{eq:cwf}) and the ``effective bound state WF''
\begin{equation}
r^{-1}\psi_b(r)[Y_1(\hat{r})\chi_{1/2}]_{J=3/2}
\end{equation}
obtained from Eq. (\ref{eq:gswf}):
\begin{equation}
r^{-1}\psi_b(r)=\int_0^{\infty}\phi_{011}(\rho ,r)\rho d\rho
\phi^{^8{\rm Be}}(\rho).  \label{eq:eff}
\end{equation}

Using Eqs.~(\ref{eq:cwf}), (\ref{eq:eff}) the cross section
is of the same form as in the single--particle case. The
total (s--wave) photodisintegration cross section
is calculated in the simple form
\begin{eqnarray}
\sigma=(2/3)^6\pi(e^2/\hbar c)(2\mu/\hbar^2)E_{\gamma}
k^{-1}I^2,    \label{eq:crs} \\
\mbox{\rm where~~~~~} 
I=\int_0^{\infty}\psi_{\rm b}(r)\psi_{\rm sc}(r)rdr.  \label{eq:int}
\end{eqnarray}
Here $\mu$ is the $n^8$Be reduced mass, and $(\hbar k)^2/(2\mu)$ is
the excitation energy $E_{\gamma}-E_{\rm th}$ that will be denoted as $E$
below. In case of a single--particle description of the process,
i.e., the ``valence neutron'' model, Eq.~(\ref{eq:crs}) is valid with the
bound--state WF
normalized to unity, while in our case (see below)
\begin{equation}
\int_0^{\infty} \psi_{\rm b}^2(r)dr=0.43. \label{eq:red}
\end{equation}
It is implied here and in Eq. (\ref{eq:crs}) that
the $^8$Be and $^9$Be ground state wave functions
are normalized to unity. The function $\psi_{\rm b}(r)$ is shown in Fig.~2
(dashed line).

\section{The photodisintegration cross section}

In Ref.~\cite{corman63} the data at the energies up
to 185\,keV above threshold ($E_{\gamma}\le 1.85$ MeV) 
from Refs.~\cite{gib59,john62} were reproduced at
a qualitative level within the following framework. The 
valence neutron model with
Woods--Saxon type potentials was used. The results obtained in this way 
were multiplied by some constant factors less than unity,
so called ``reduction factors'', to approach the experimental cross section.
To obtain the continuum wave function
the depth of the potential was varied while the radius
and diffuseness parameters were taken the same as for the central
component of the potential in the bound state calculation.
Two fits were found, one with a reduction factor of 0.53 leading
to a weakly bound $^8$Be$+n$ $1/2^+$ state, and the other one with
a reduction factor of 0.31 leading to a virtual state. The first
possibility was once preferred in view of the results of
Ref. \cite{blair61}. In that work a two channel $^8$Be$+n$ model
of the ground state of $^9$Be was introduced to cure the
single--particle description \cite{francis60} of 
photodisintegration. The $^8$Be subsystem was allowed to
be in the ground and first excited state, and that led to the
reduction factor of 0.5 or 0.6 in the cross section
depending on the assumptions. In Ref. \cite{barker61} the
reduction factor of 0.56 was found for that model. In contrast
to Ref. \cite{corman63} our results below definitely testify
to a virtual $^8$Be$+n$ $1/2^+$ state. This is probably due
to a more realistic treatment of the ground state of $^9$Be
in our model.
%a more flexible and realistic ground state WF of the $^9$Be
%nucleus we use. Another point is that the range of the $^8$Be$+n$
%potential in the excited state should not be constrained to
%the ground state.
In fact the reduction factors obtained in the two--channel model of 
the ground state of $^9$Be \cite{blair61}
should be used in conjunction with the channel coupling n$+^8$Be
dynamics, instead of
using \cite{corman63} single--channel dynamics.
%In fact the simple potential model description
%of the
%ground state \cite{corman63} is inconsistent with
%reduction
%factors that imply coupling channel potentials also to be included.

In Ref. \cite{mahaux65} the same data were fitted with the line
shape $\sqrt{E}(E+\bar{E})^{-2}$. This shape was derived incorrectly 
from the Breit--Wigner cross section under the assumption that the
$1/2^+$ level of $^9$Be is a bound or virtual $^8$Be$+n$ state.
%One can see that the true line shape under this assumption is
%$\sqrt{E}[(E+\bar{E})(E+E_1)]^{-1}$ with $E_1\gg \bar{E}$.
In Ref.~\cite{barker83} the data of
Ref.~\cite{fui82} were fitted with a one--level R--matrix
approximation. The fit leads to a complex--energy resonant
state \cite{barker83} and the real part of the
complex energy proves to be negative.

First we shall analyze the data of Refs. \cite{gib59,john62,ham53}
(full circles, full squares, and full diamond in Fig. 1).
A search of the parameters $V_0$, $r$, and $a$ of the potential
(\ref{eq:ws}) giving an acceptable fit to the data is performed. 
Several local minima of the quantity $\chi^2$ in the space of the
parameters are found. One of them is provided by
\begin{equation}
V_0=35.99\,{\rm MeV},\,\,\,\,\,\,\,\,\,R=3.126\,{\rm fm},
\,\,\,\,\,\,\,\,\,a=0.8108\,{\rm fm}.   \label{eq:1}
\end{equation}
These values seem to be very reasonable. For this set the
$X=\chi^2/$(degrees of freedom) value equals 0.62. Another one is
obtained with the parameters
\begin{equation}
V_0=52.86\,{\rm MeV},\,\,\,\,\,\,\,\,\,R=2.006\,{\rm fm},
\,\,\,\,\,\,\,\,\,a=1.051\,{\rm fm}   \label{eq:2}
\end{equation}
giving $X=0.525$. Several other minima also exist with sizable
higher but still acceptable $X$ values. In Fig.~3 the
photodisintegration cross sections obtained with the parameters
(\ref{eq:1}) and (\ref{eq:2}) are shown as the solid and dashed curve,
respectively. 
The two cross sections prove to be quite close to each other. To
clarify partially the reason for this we note that the biggest contribution
to the matrix element of Eq.~(\ref{eq:int}) comes from the distances
beyond the range
of the potential. The distances larger than 5 fm in Eq.~(\ref{eq:int})
provide 60-70\% contribution to the cross section, and at such distances 
the wave functions deviate  from the asymptotic ones,
Eq.~(\ref{eq:sin}), by not more than 10\% (except for regions in
the vicinity of zeros).
The asymptotic wave functions are determined by the phase shifts
i.e., predominantly by the scattering length
$a$ and the effective range $r_0$. Therefore
%Let us take $r_{\rm cut}=3.6$\,fm instead of zero 
%as the lower limit of integration in Eq.~(\ref{eq:int}).
%Then the cross sections
%will change not more than by 2\% for both potential sets (\ref{eq:1}) and (\ref{eq:2}) 
%for any energy.
%But for $r\ge r_{\rm cut}$
%the functions $\psi_{\rm sc}(r)$ are mainly determined by the phase shifts
%entering Eq.~(\ref{eq:sin}), i.e., predominantly by the scattering length
%$a$ and the effective range $r_0$. Therefore  
the procedure is equivalent to some degree to fitting $a$ and $r_0$
values. Once this is done,
the cross sections are not very
dependent on the particular set of the
potential parameters. 
The $a$ and $r_0$ values are $-27.6$\,fm and $8.79$\,fm, respectively, for the
set (\ref{eq:1}), and $-28.4$\,fm and
$9.77$\,fm  for the set (\ref{eq:2}). 
All the other above mentioned sets of potential parameters 
giving local minima to $X$ lead to
very similar $a$ and $r_0$ values. However considerable changes in 
the scattering WF inside the potential can influence the results,
see the next paragraph.
%In view what was said above, it appears that
%the effective range value overestimates the real range of the potential in our case. 
%This overestimation can be understood as follows. Provided that the scattering length is large
%the effective range proves to be a good
%measure of the range of the potential when the zero energy scattering wave function does not 
%oscillate inside the well.
%This condition is satisfied in the NN scattering case, and it means that the 
%potential does not support deeply bound states: only one weakly bound state exists or 
%close to appear. But in addition to the virtual level our potentials support one more
%bound state that is natural for the $s$--neutron case from the shell--model
%point of view. Due to this fact the zero energy wave function crosses zero inside the well 
%that increases
%the mean deviation of its square from the square of the asymptotic wave function.
%This leads to an increase in the effective range value.

The following way to interpolate between the data has also 
been tried. 
Let us denote by
$\sigma_0(E)$ the cross section obtained in case when  the 
right--hand side of Eq.~(\ref{eq:sin}) is used as a continuum 
wave function
for all $r$ values. This cross section  has been calculated taking 
$\delta$ from the effective range expansion with 
the $a$ and $r_0$
values given by the potential (\ref{eq:1}). Let us represent  
$\sigma$ as 
$c(E)\sigma_0(E)$ and fit $c(E)$ to experiment. It is hoped that, in contrast 
to $\sigma$, the
factor $c(E)$ behaves in a smooth way and thus can be reliably obtained
from an interpolation procedure. Indeed, the behavior of both
$\sigma$ and $\sigma_0$  can be approximately described by the resonant factor
$k^{-1}\sin^2\delta$ times a slowly varying function. Even a fit with  
$c(E)=const=0.55$ proves to provide a sufficiently low $X$ value. The cross 
section obtained with this $c$ is shown in Fig.~3 as the dotted curve. 
Presumably this procedure provides less accurate results than the previous one.
Of course, the energy
dependence $k^{-1}\sin^2\delta$ is not accurate enough in the whole energy 
range, as a comparison with the exact solution for the potential (\ref{eq:1})
shows. Hence $c(E)$ should include an energy dependence, but this could not
be determined because of experimental uncertainties.    

Next we applied the procedure to the data of Ref.~\cite{fui82} (open circles
in Fig. 1). 
Three local minima with acceptable $X$ values are found. However the 
parameters of the potential corresponding to all of them: 
\begin{eqnarray}                       
V_0=11.0\,{\rm MeV},\,\,\,\,\,\,R=2.35\,{\rm fm}\,\,\,\,\,a=0.258\,{\rm fm}
\nonumber \\
V_0=6.49\,{\rm MeV},\,\,\,\,\,\,R=3.10\,{\rm fm}\,\,\,\,\,a=0.260\,{\rm fm}
\label{fu}\\
V_0=15.8\,{\rm MeV},\,\,\,\,\,\,R=1.37\,{\rm fm}\,\,\,\,\,a=0.958\,{\rm fm} 
\nonumber\end{eqnarray}
prove to be rather unrealistic. There exists one more difference between these
potentials and those in Eqs. (\ref{eq:1}) and (\ref{eq:2}). The latter 
potentials, as well as the other potentials (\ref{eq:ws}) providing a good fit
to the same data, 
support one deeply bound state and one state close to being bound. On the
contrary, 
the potentials (\ref{fu}) support only one very weakly bound state. 
An existence of one deeply bound $s$ level in the neutron mean field in the
$^9$Be nucleus, or, equivalently, one node in the low--energy scattering wave function inside 
the potential, seems to be natural from the shell--model point of
view. We think this point of view is sufficient to establish the correct number
of nodes for the neutron motion inside the Woods--Saxon potential ,
even for such a clusterized system. In the $\alpha$--particle oscillator
model of $^9$Be \cite{kunz60}, for example, the first allowed neutron $s$--state 
contains a substantial admixture of the nodeless $0s$ function, but this 
leads not to a disappearance but only to a shift of the node. If one admits that 
the state considered is a mixture of $0s$ and $1s$ oscillator functions then
there exists
just one node located within the distance of 3 fm from the origin.
Therefore we conclude that in the region where various data sets differ from 
each other the older
radioactive isotope data are preferable. We also note that the cross
section we obtain with potentials (\ref{eq:1}) and (\ref{eq:2}) for the highest energies 
considered, being lower than 
the fitted datum of Ref.~\cite{ham53}, agrees well with the  
bremsstrahlung Jacobson data \cite{jacob61}.  
  
\section{The astrophysical reaction rates}

The $^9$Be$+\gamma\rightarrow \alpha+\alpha+n$ reaction rate per nucleus 
per time unit is calculated via the usual averaging
the elementary photodisintegration cross section $\sigma(E_\gamma)\cdot c$ with the
approximate, or Wien distribution for the photon density,
\begin{equation}
\lambda_\gamma=c\pi^{-2}(\hbar c)^{-3}\int_{E_{\rm th}(^8{\rm Be})}^\infty\sigma(E_\gamma) 
E_\gamma^2\exp(-E_\gamma/kT)dE_\gamma,  \label{lam}
\end{equation}
where $E_{\rm th}(^8{\rm Be})=1.6654$ MeV. The rate of the reverse reaction (the number
of reactions per time unit per unit volume) is 
%given by the relation
%\cite{fowler67} 
\begin{equation}
P(\alpha\alpha n)=(1/2)n_\alpha^2n_n\langle\alpha\alpha n\rangle,
\end{equation}
where $n_\alpha$ and $n_n$ are numbers of particles per unit volume. The 
reaction 
constant $\langle\alpha\alpha n\rangle$ is obtained from Eq. (\ref{lam})
using the reverse ratio $RR$ \cite{fowler75,fowler67}:
\begin{eqnarray} 
N_{\rm A}^2\langle\alpha\alpha n\rangle&=&\lambda_\gamma/RR, \label{3} \\
RR&=&5.84\cdot10^{19}T_9^3\exp(-18.261/T^9).
\end{eqnarray}    
Here $N_{\rm A}$ is Avogadro's number, $T_9$ is the temperature in $10^9$ K, 
$18.261/T^9=E_{\rm th}(\alpha\alpha n)/kT$ with $E_{\rm th}(\alpha\alpha n)=1.5736$ MeV,  
and it is implied that the 
quantities (\ref{lam}) and (\ref{3}) are given in sec$^{-1}$ and cm$^6$ 
sec$^{-1}$
mole$^{-2}$, respectively.
%In Eq.~(\ref{lam}) the integration is performed up to $E_\gamma=12$\,MeV
%that suffices for this range of temperatures. 
Use of the Wien distribution
instead of the exact, or Planck, one, i.e.
$[\exp(E_\gamma/kT)-1]^{-1}\rightarrow\exp(-E_\gamma/kT)$, allows 
application of the above listed simple reverse ratio theory. For temperatures of
$T_9=5$ and 10, for example, it gives the reaction constant (\ref{lam}) with relative errors of 1\,\%
and 5.4\,\%, respectively.
For $E_{\gamma}\le 2.2$\,MeV the cross section 
$\sigma(E_\gamma)$ obtained in the 
preceding 
section with the potential (\ref{eq:1}) is used. For $E_\gamma$ from 2.2\,MeV 
up to
5\,MeV the Jacobson bremsstrahlung data \cite{jacob61} are used. The former energy region
provides 96\% and 62\% contribution to the cross section for $T_9=2$ and 5, respectively.
The contribution from energies $E_\gamma$ higher than 5\,MeV reaches 0.3\% and 6.7\%
for $T_9=5$ and 8, respectively. 
%Between the minimum at 2.2\,MeV and the maximum at 2.45\,MeV the Jacobson data
%were increased by $\simeq 4$\% to match our low--energy cross section.  
%The data from \cite{bert60} and \cite{edge56} 
%for $5\le E_\gamma\le 7$\,MeV and those  
%from \cite{costa66} and 
%\cite{nat53} for 7 $\le E_\gamma\le$ 12\,MeV are used. 

The values of the rate (\ref{3}) obtained can be represented by the fit
\begin{equation}
N_A^2\langle\alpha\alpha n\rangle=6.59\times10^{-6}T_9^{-3/2}\exp[-(1.0653/T_9)]
\left(1+\sum_{n=1}^7a_nT_9^n\right)^{-1} \label{fit}
\end{equation}
with 
\begin{eqnarray*}
a_1 & = & 3.3562\,\,\,\,\,a_2=-0.86389\,\,\,\,\,a_3=0.42268\,\,\,\,\,
a_4 = -0.14913\\
a_5 & = & 2.7039\times10^{-2}\,\,\,\,\,a_6=-2.4000\times10^{-3}
\,\,\,\,\,a_7=8.3223\times10^{-5}
\end{eqnarray*}
The fit reproduces our $N_A^2\langle\alpha\alpha n\rangle$ values with the accuracy
better than 1\% at any $T_9$ in the range $0.1\le T_9\le 8$. In Eq. (\ref{fit}) 
$1.0653/T_9=[E_{\rm th}(^8{\rm Be})-E_{\rm th}(\alpha\alpha n)]/kT$. The factor $T_9^{-3/2}\exp[-(1.0653/T_9)]$
represents the asymptotic behavior of the $^8$Be formation contribution to 
the rate when $T_9$ tends to zero. 

In the table our values for the three--body reaction rate (\ref{3}) are
compared with those of Ref.~\cite {fowler75} and those of Ref.~\cite{gor95}
where constructive or destructive interference between
the resonant and non--resonant contributions at energies above
the resonance energy was assumed.

Summarizing, we have constructed a semi--microscopic model for the low--energy
photodisintegration
of the $^9$Be nucleus and have analyzed the experimental data with its help.
Our analysis supports 
the older radioactive isotope data. The theoretical cross section we derived may
be compared
with future microscopic calculations of the process. We have calculated the
astrophysical rates
for the reaction $\alpha+\alpha+n\rightarrow ^9$Be$+\gamma$ and the reverse
reaction. Our new reaction rates agree at $T_{9} =2.0$ with the ones given
in Ref.~\cite{fowler75}. They are somewhat smaller (larger) for lower
(higher) temperatures than $T_{9} =2.0$. The reaction rates
given in Ref.~\cite{gor95}
agree much better with our reaction rate at higher temperatures if
one assumes
in Ref.~\cite{gor95} constructive (destructive) interference
between the resonant and non--resonant contributions at energies above
(below) the resonance energy.
 
We are indebted to J.S. Vaagen and J.M. Bang for very fruitful comments. 
This work was supported partially by the the Fonds
zur F\"orderung wissenschaftlichen Forschung in
\"Osterreich (project P10361--PHY) and the Russian Foundation for Basic Research (grant no 
97-02-17003).
\medskip
 
\section*{APPENDIX}

In our calculation
of the ground state of $^9$Be the $\alpha\alpha$ potential
 is taken in the form \cite{ali66}
\[ V_R\exp[-(\mu_R\rho)^2]-V_A\exp[-(\mu_A\rho)^2] \]
with $V_A=130$\,MeV, $\mu_A=0.475$\,fm$^{-1}$, $\mu_R=0.7$\,fm$^{-1}$,
and $V_R=500$, 320, and 10\,MeV for $l$=0, 2, and 4, respectively.
The Coulomb $\alpha\alpha$ interaction is also added. The
$n\alpha$--
interaction in s--, p--, and d--states is taken into account.
As in many previous studies \cite{zhu93} the s--wave repulsive potential
$V\exp[-(r/R)^2]$ with $V=50$\,MeV and $R=2.3$\,fm is used. The
initial potential in p-- and d--states \cite{bang} includes
central and spin--orbit components:
\[ V(r)=-V\left (1+e^{\frac{r-R}{a}}\right )^{-1}-U ~ {\bf l} \cdot {\bf s} ~
\frac{1}{r}\frac{d}{dr}\left (1+e^{\frac{r-R_1}{a_1}}\right )^{-1} \]
with $V=43$\,MeV, $R=2$\,fm, $a=0.7$\,fm, $U=40$\,MeV\,fm$^2$,
$R_1=1.5$\,fm, and $a_1=$0.35\,fm. The parameter $V$ is reduced to 39.6\,MeV
in the present three--body calculation, in order to reproduce the empirical g.s. energy.

The three--body dynamic
equation is written in the form of the Faddeev differential
equations and each Faddeev component is expanded over
hyperspherical harmonics and hyperradial basis functions.
Using the Raynal--Revai rotations
of hyperspherical harmonics\footnote{A review of the hyperspherical
formalism can be found e.g. in Ref. \cite{zhu93}.}
the matrix elements are reduced
analytically to two--dimensional integrals. 
The equations are projected onto subspaces of the basis
functions retained that reduces the problem to the algebraic eigenvalue problem. 
The number of basis functions retained is quite high and ensures the
adequate convergence of the calculation.

\begin{table}[tb]
\begin{tabular}{ccccc}
 %\hline
\multicolumn{1}{l}{$T$ [$10^9$\,K]}&
\multicolumn{4}{c}{$N_A^2\langle\alpha\alpha n\rangle$}\\\hline
& Present work & Ref.~\cite{fowler75} & Ref.~\cite{gor95} (destructive)
& Ref.~\cite{gor95} (constructive)\\
\hline
0.2&0.21$\cdot10^{-6}$&0.30$\cdot10^{-6}$&---&---\\
0.5&0.87$\cdot10^{-6}$&1.1$\cdot10^{-6}$&$0.55\cdot10^{-6}$&$0.61\cdot10^{-6}$\\
1.0&0.60$\cdot10^{-6}$&0.67$\cdot10^{-6}$&$0.32\cdot10^{-6}$&$0.44\cdot10^{-6}$\\
2.0&0.23$\cdot10^{-6}$&0.23$\cdot10^{-6}$&$0.12\cdot10^{-6}$&$0.20\cdot10^{-6}$\\
3.0&0.12$\cdot10^{-6}$&0.99$\cdot10^{-7}$&$0.60\cdot10^{-7}$&$0.11\cdot10^{-6}$\\
4.0&0.73$\cdot10^{-7}$&0.52$\cdot10^{-7}$&---&---\\
5.0&0.51$\cdot10^{-7}$&0.31$\cdot10^{-7}$&$0.26\cdot10^{-7}$&$0.52\cdot10^{-7}$\\
%10&0.21$\cdot10^{-7}$&0.53$\cdot10^{-8}$\\
%15&0.16$\cdot10^{-7}$&0.17$\cdot10^{-8}$\\\hline
\end{tabular}
\end{table}
\begin{figure}
\caption{Experimental data on the low--energy cross section for 
photodisintegration of $^9$Be. 
Bremsstrahlung data: Ref. [12] (solid curve representing the fit of the authors 
to their data), Ref. [11] (star representing
the maximum of the spectrum; the rest of the spectrum is not shown).
Radioactive isotope data: Ref. [13] (full circles), Ref. [14] (full squares),
Ref. [15] (full diamond), Ref. [16] (open circles)}
%e 
%del (see text).}
\end{figure} 

\begin{figure}
\caption{The wave function of the $^8$Be resonance (solid curve) and
the ``effective'' n--$^8$Be relative motion wave function in $^9$Be
(dashed curve). The former wave function is constructed as the $s$--wave
continuum solution in the $\alpha\alpha$ potential (see 
appendix) at the energy of 0.09518 MeV, the peak of the
resonance in this potential.} \end{figure}

\begin{figure}
\caption{Calculated cross sections for the photodisintegration of $^9$Be.
Solid and dashed curves are obtained with the parameters (10) and (11),
respectively. Dotted curve corresponds to the alternative method listed
in the paragraph after Eq. (11). The experimental data are as in Fig. 1.}
\end{figure}
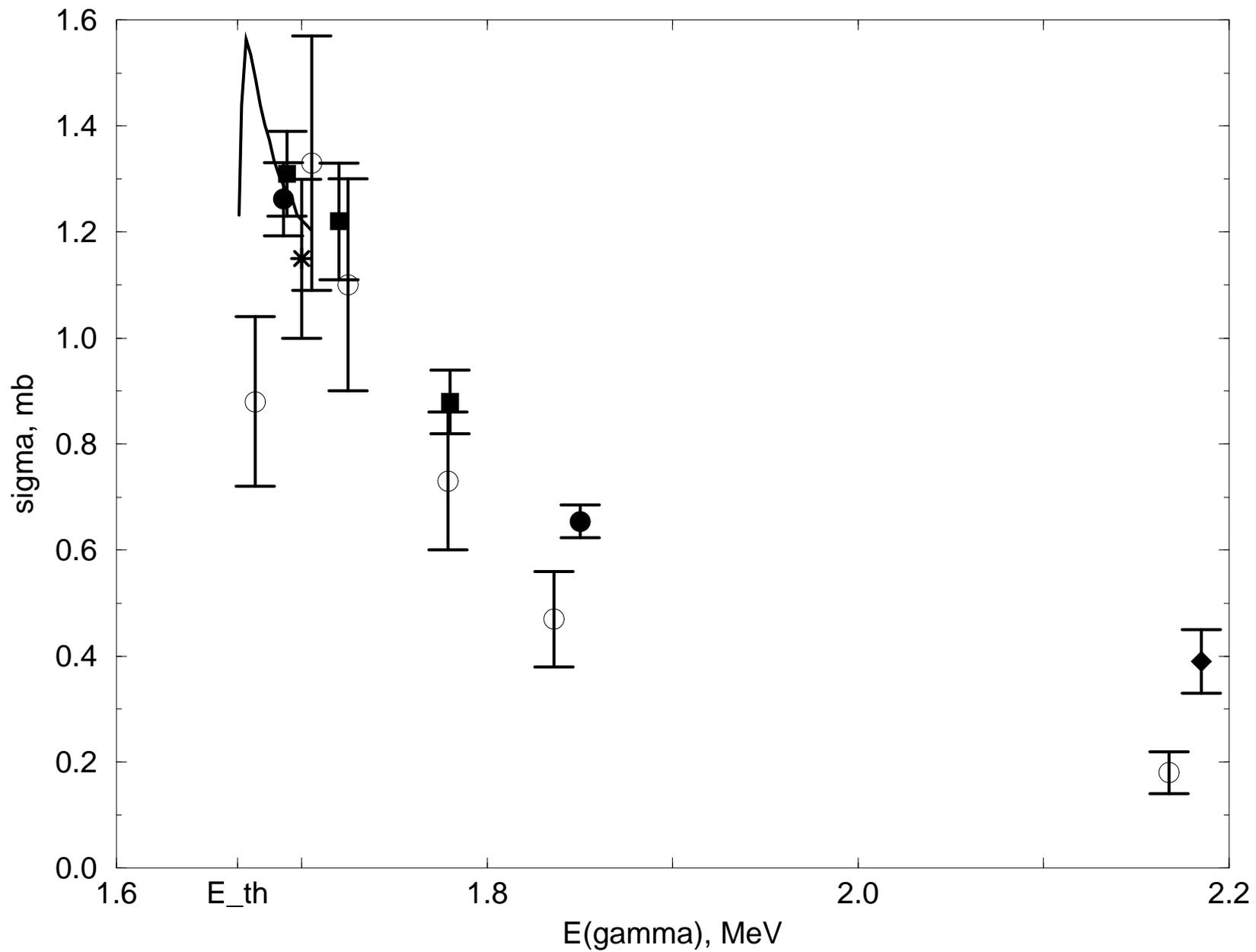
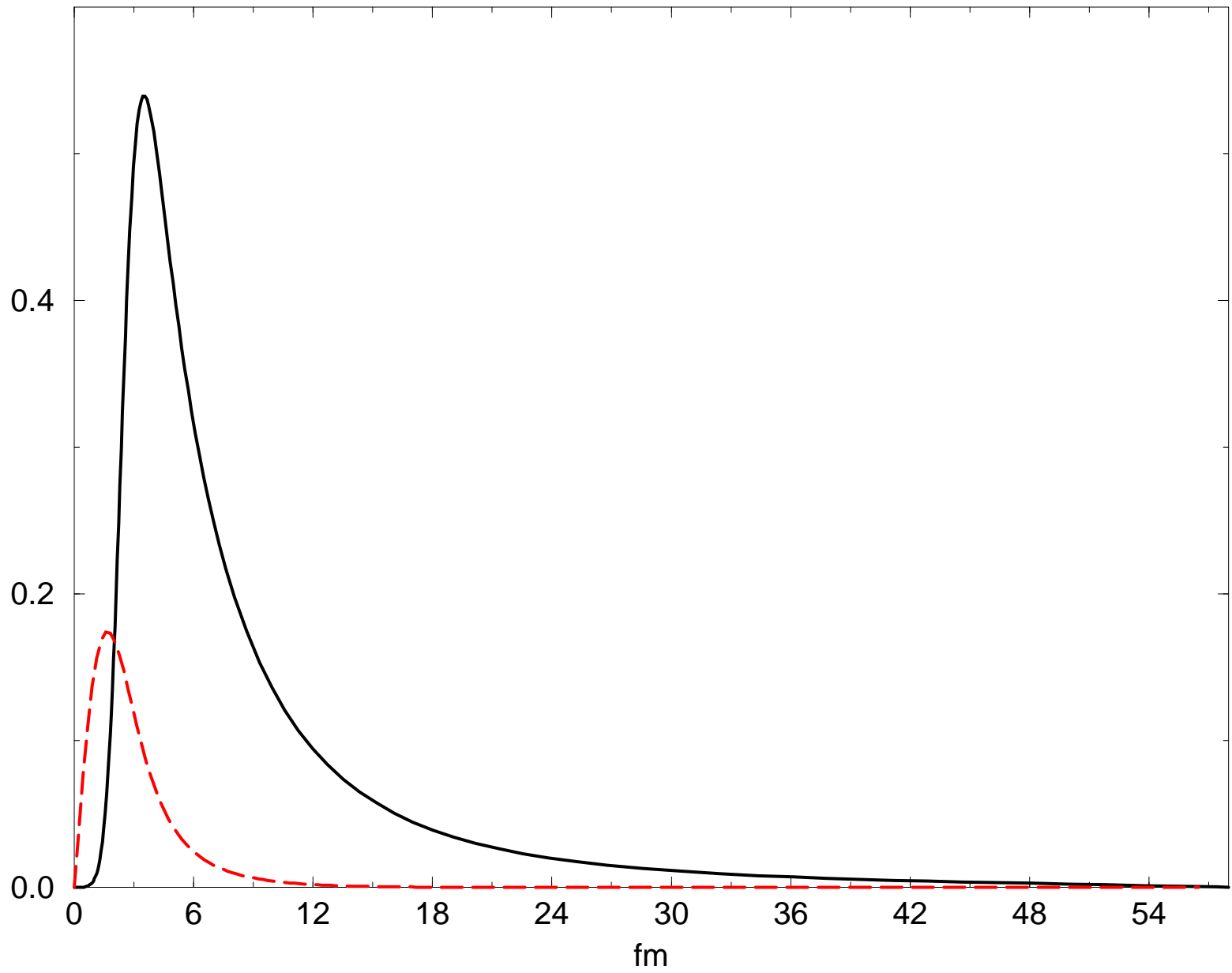
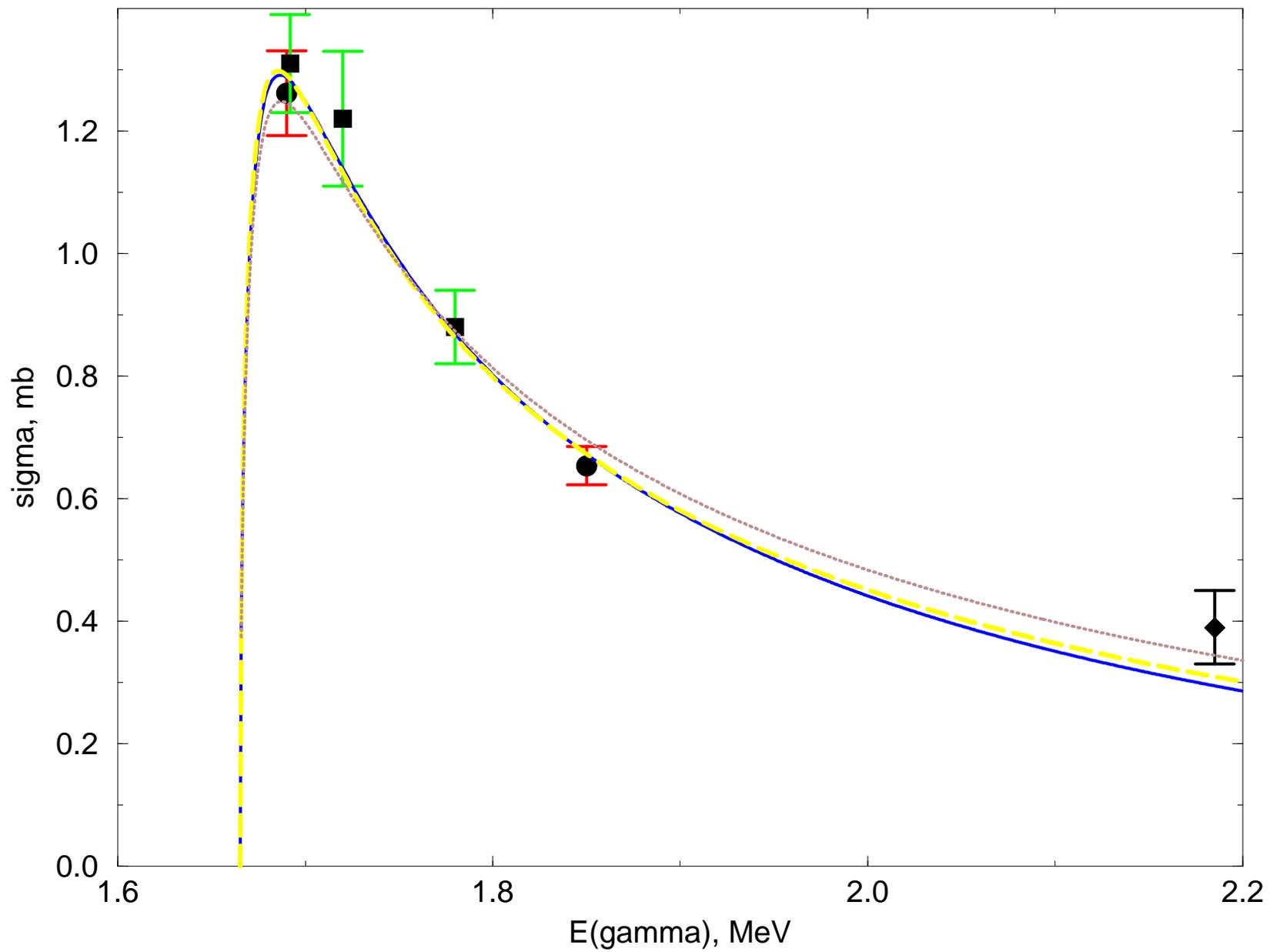
\end{document}